# Dynamics of Two Distinct Exciton Populations in Methyl-functionalized Germanane

*Eugenio Cinquanta[1*], Samim Sardar[2], Warren L. B. Huey[3], Caterina Vozzi[1], Joshua E. Goldberger[3], Cosimo D'Andrea[2,4*], and Christoph Gadermaier[2,4*]*

1) Istituto di Fotonica e Nanotecnologie, Consiglio Nazionale delle Ricerche, piazza L. da Vinci 32, 20133 Milano, Italy;
2) Center for Nano Science and Technology @PoliMi, Istituto Italiano di Tecnologia, Via G. Pascoli 70, 20133 Milan, Italy ;
3) Department of Chemistry and Biochemistry, The Ohio State University, Columbus, Ohio 43210, United States;
4) Dipartimento di Fisica, Politecnico di Milano, piazza L. da Vinci 32, 20133 Milano, Italy;

**Corresponding Author:**

*eugenioluigi.cinquanta@cnr.it*

*cosimo.dandrea@polimi.it*

*christoph.gadermaier@polimi.it*

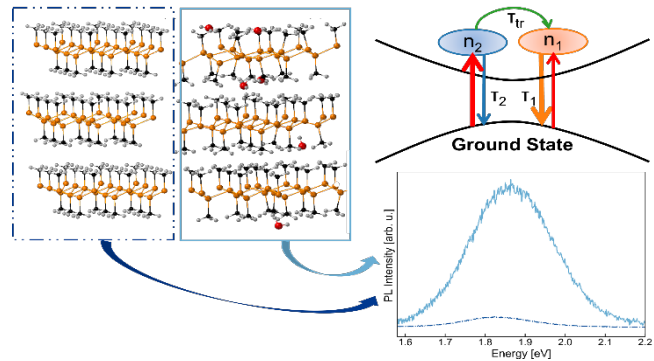

## Abstract

Methyl-substituted Germanane is an emerging material that has been proposed for novel applications in optoelectronics, photoelectrocatalysis, and biosensors. It is a two-dimensional semiconductor with a strong above-gap fluorescence associated with water intercalation. Here, we use time-resolved photoluminescence spectroscopy to understand the mechanism causing this fluorescence. We show that it originates from two distinct exciton populations. Both populations recombine exponentially, accompanied by the thermally activated transfer of exciton population from the shorter- to the longer-lived type. The two exciton populations involve different electronic levels and couple to different phonons. The longer-lived type of exciton migrates within the disordered energy landscape of localized recombination centers. These outcomes shed light on the fundamental optical and electronic properties of functionalized germanane, enabling the groundwork for future applications in optoelectronics, light-harvesting, and sensing.

Two-dimensional (2D) materials are presently one of the most actively explored platforms for the development of nanoscaled (opto)electronic devices [1 - 3]. Monoelemental 2D materials (Xenes) and their substituted counterparts (Xanes, e.g., GeH or $GeCH_3$) are rapidly emerging alongside the much more well-studied transition metal dichalcogenide semiconductors due to high electron mobility, a wide range of band gaps, and the possible tuning of their morphology and physical properties [4 - 10]. Germanane has been proposed recently as a novel active material for optoelectronics, photoelectrocatalysis, anti-bacterial coating, and biosensors, with the specific performances determined by the functional groups [11 - 16].

The photoluminescence (PL) of multilayer $GeCH_3$, conversely to its H- terminated counterpart, is tightly linked to the presence of water in the Van der Waals gap [17]. Water intercalation switches the PL spectrum reversibly between a bright red peak centered around 1.97 eV – significantly above the 1.62 eV bandgap – for the hydrated material, and a broad band-tail emission for the dry one. The PL excitation spectrum of the 1.97 eV emission starts at 2.1 eV and has its maximum at 3.5 eV, hence demonstrating that this emission arises from strong electronic transitions involving electronic levels above the conduction band minimum and/or below the valence band maximum. The strong above-gap PL and simultaneous suppression of the band-tail emission suggest that the involved above-gap levels have no allowed relaxation channel towards the band edges. A deeper insight into the electronic nature of the involved excited states, the interplay between them, and the associated time scales [18-20] are vital for rationalizing Xanes' optoelectronic and light-harvesting functionalities.

In this work, we exploit time-resolved photoluminescence (TRPL) to unveil the origin of the 1.97 eV above bandgap emission in $GeCH_3$. From the analysis of the emission peak energy and intensity as a function of time and temperature, we assign the observed fluorescence to the interplay of two distinct exciton populations and discuss their electronic nature.

The polycrystalline powders of $GeCH_3$ were synthesized following previously established procedures [17].

For the TRPL measurements, we used a Ti:Sapphire oscillator (Chameleon Ultra II Coherent) producing a train of 140-fs pulses with a repetition rate of 80 MHz at 800 nm. A β-barium borate (BBO) crystal was used to obtain the second harmonic at 400 nm. Spatial resolution was achieved through the incorporation of a homemade microscope in the setup [21]. A long pass dichroic mirror at 530 nm was used to reflect the excitation beam (400 nm) that was then coupled to a 20X objective (Nikon) to focus onto the sample with a spot size of about 6 μm. The emission signals were collected in back-scattering geometry using a 550 nm long-pass filter and analyzed by a spectrograph (Princeton Instruments Acton SP2300) coupled to a streak camera (Hamamatsu C5680, Japan) equipped with a synchro-scan voltage sweep module. In these measurements, the fluorescence intensity was obtained as a function of both wavelength and time with spectral and temporal resolutions of ~1 nm (~3 meV in our spectral range) and ~20 ps (for 2 ns time window), respectively. Cryogenic measurements were performed using a cryostat (Oxford Instruments) cooled with liquid nitrogen under vacuum conditions ($10^{-6}$ mbar). From the synthesized powder, we selected a bulk flake of a few hundred-micron lateral sizes and glued it with two thin slices of carbon tape onto a fused silica substrate. The sample was then gently annealed at $150^0C$ under vacuum for 15-18 hours before the measurements to remove most of the intercalated water and reach a water concentration that remains stable during the measurements at varying temperatures.

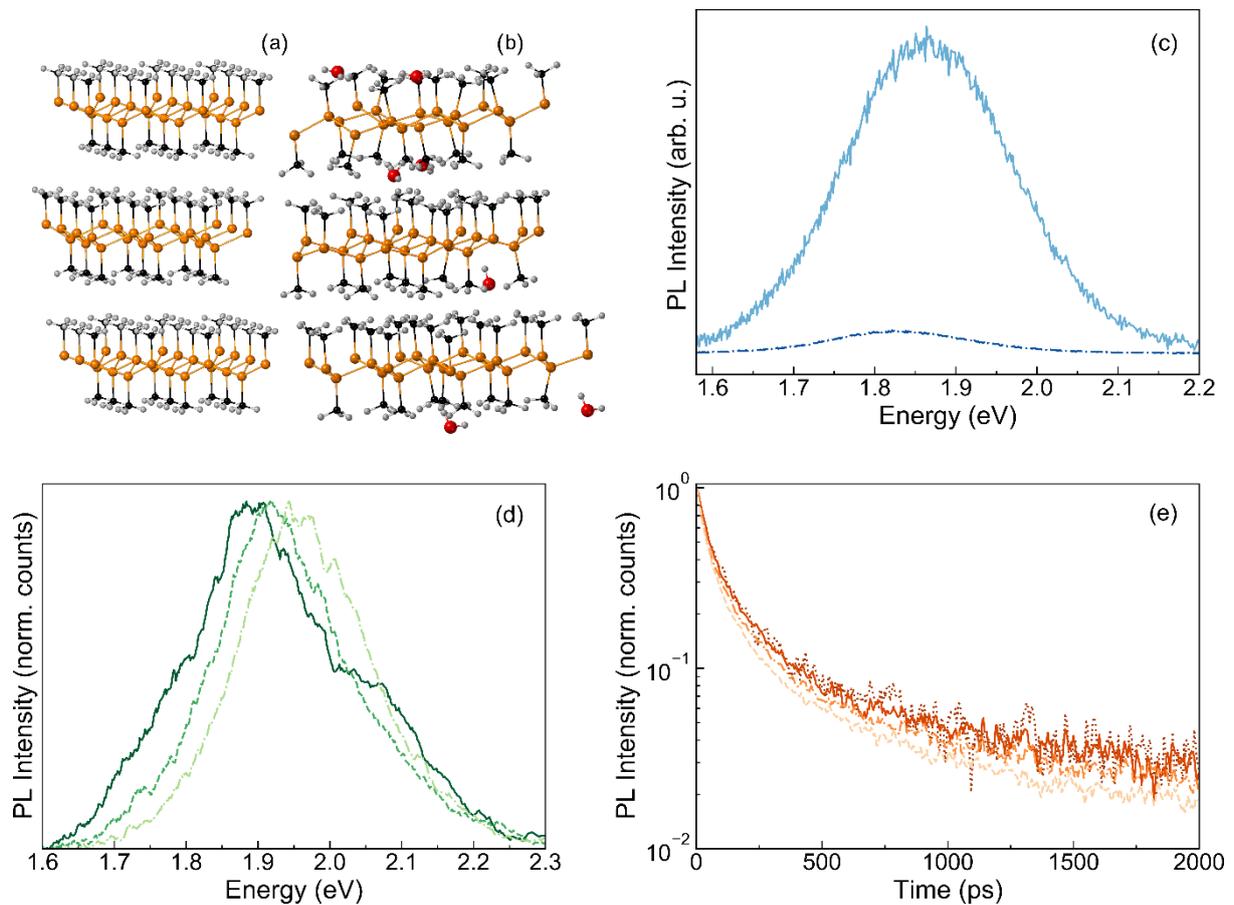

*Figure 1. Atomic structure of GeCH$_3$ a) without water intercalation and b) with water intercalation. Ge atoms are orange, Carbon atoms are black, hydrogen atoms are white, and oxygen atoms are red. c) Photoluminescence integrated over time for dry (under vacuum, dash-dotted blue curve) and hydrated (in air, turquoise solid curve ) GeCH$_3$ acquired with a mean excitation power of 15 µW; d) normalized TRPL spectra of the vacuum-treated "dry" sample at 77 K integrated into the 0<t<100ps (dash-dotted curve), 200ps<t<300ps (dashed curve) and 900<t<1000ps (solid curve) temporal windows; e) normalized spectrally integrated (1.65 – 2.25 eV) dynamics of the vacuum-treated "dry" sample at 77K upon mean excitation powers of 3µW (dots), 10 µW (solid curve), 30 µW (dash-dotted curve) and 100 µW (dashed curve).*

Figures 1a and 1b show the ball and stick model of dry and water intercalated bulk GeCH$_3$. For the investigation of the above bandgap emission, TRPL characterization was performed as described above. Figure 1c shows the time-integrated (0-2ns) TRPL spectra for the dry (dash-dotted blue curve) and the hydrated (solid turquoise curve) sample. The fluorescence is largely quenched but still clearly detectable when the sample is placed in 10$^{-6}$ mbar, hence confirming that the remaining $H_2O$ molecules are enough to induce the 1.9 eV emission [17].

The fluorescence of the vacuum-treated "dry" sample integrated into different temporal windows ($0<t<100ps$, $200ps<t<300ps$ and $900<t<1000ps$), shown in Figure 1d, changes shape and peak position with time $t$ after excitation, indicating that the fluorescence originates from more than just one population of recombining e-h pairs. In this respect, previous TRPL measurements on $GeCH_3$ flakes revealed the presence of two emitting species, therein assigned to mid-gap trap states and the band tail emission [22]. The PL traces integrated over the spectral range 1.65 - 2.25 eV, as shown in Figure 1e, decay almost independently of the excitation fluence over two orders of magnitude. We deduce that the relevant relaxation processes are linear with the density of photogenerated population $n$, i.e., follow an exponential decay and do not involve any interaction between non-geminate photoexcited species. Indeed, if free charge carriers were photogenerated, we would expect a more noticeable change in the recombination dynamics with increasing fluence, due to the rate proportional to $n^2$ of such dynamics [23]. Therefore, our observation is consistent with the prediction of excitons with hundreds of meV binding energy as the primary photoexcited species in Germanane [22, 23], as observed in other 2D semiconductors [26 - 29].

To gain further insight into the electronic nature and the recombination dynamics of the emitting states we explored the temperature dependence of TRPL from 77 K to 323 K. Figures 2a and 2b show the spectrally integrated PL traces. Remarkably, the dynamics depend non-monotonically on temperature. In the range from 77 K to approximately 200 K, the dynamics become gradually slower with increasing temperature, while at higher temperatures they quickly become faster again.

We propose a simple model for the temporal evolution of the fluorescence, sketched in the inset of Figure 2c and formulated in terms of rate equations (Eq. (1)). We assume two distinct exciton populations $n_1$ and $n_2$, both localized at water-induced recombination centers

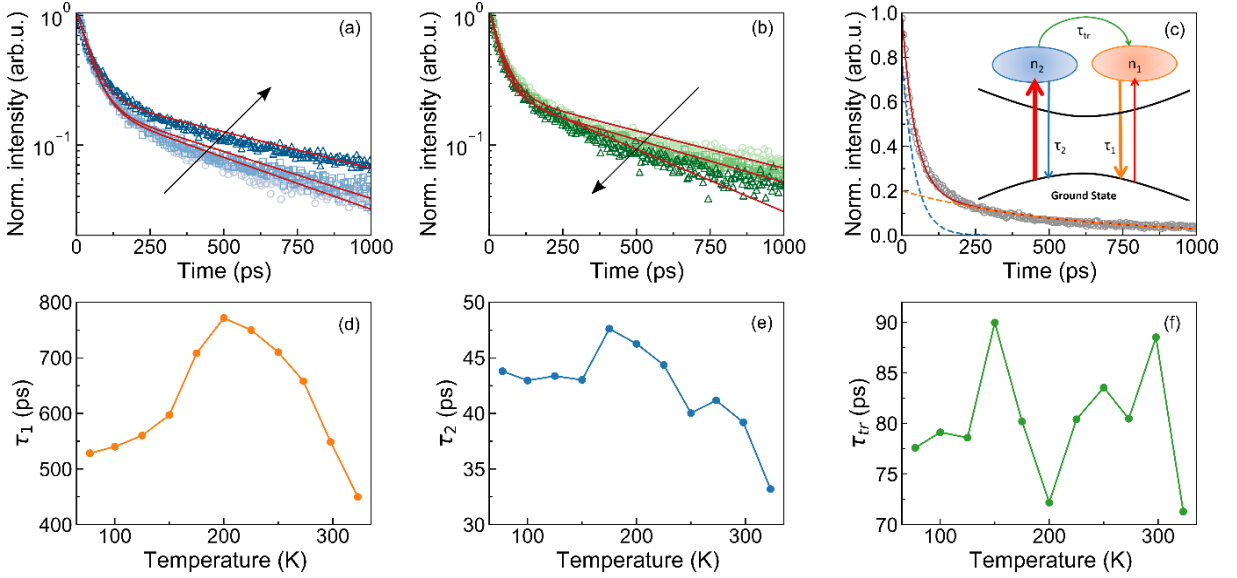

*Figure 2 a) spectrally integrated dynamics acquired upon 10 µW excitation power at 77K (circles), 150K (squares), and 200K (triangles) together with the fit to Eq. (1) (continuous red curve); b) spectrally integrated dynamics acquired upon 10 µW excitation power at 225K (circles), 273K (squares), and 323K (triangles) together with the fit to Eq. (1) (continuous red curve). The black arrows indicate increasing temperature; c) spectrally integrated dynamics acquired at 77K (circles) upon 10 mW excitation power, fitted to Eqs (1) (continuous red curve), time-dependent populations of long- and short-lived excitons (orange and blue dashed lines, respectively); d-f) time constants $\tau_1$, $\tau_2$ and $\tau_{tr}$ extracted from the fit to Eq. (1) as a function of temperature.*

(RCs). Each population is formed at a time scale shorter than our instrument response function and decays exponentially with its own time constant $\tau_1$ and $\tau_2$, which comprise both radiative and non-radiative recombination. Additionally, we assume transfer from $n_2$ to $n_1$ with a simple, Arrhenius-like thermal activation with a prefactor $1/\tau_{tr}$ and activation energy $\Delta\varepsilon_1$ [29]:

$$\frac{dn_1}{dt} = -\frac{n_1}{\tau_1} + \frac{n_2}{\tau_{tr}} e^{-\Delta\varepsilon_1/k_b T}$$
$$\frac{dn_2}{dt} = -\frac{n_2}{\tau_2} - \frac{n_2}{\tau_{tr}} e^{-\Delta\varepsilon_1/k_b T}$$
(1)

The simple model fits the measured PL temporal evolution at different temperatures remarkably well (Figures 2a-c). For all temperatures, we obtained the best fit for initial populations of approximately 20% $n_1$ and 80% $n_2$. The exponential decay times are shown in Figures 2d and 2e, around 650 ps for $\tau_1$ and around 45 ps for $\tau_2$. These times vary by +/- 20% up to 300 K. Each of them comprises the radiative and non-radiative contributions $\tau_r$ and $\tau_{nr}$. Hence, the small variations in $\tau_1$ and $\tau_2$, which both seem to peak around 200 K, maybe due to small opposite trends of $\tau_r$ and $\tau_{nr}$ with temperature, resulting in weakly temperature-dependent PL quantum yields $\eta_1 = \tau_{nr1}/(\tau_{nr1} + \tau_{r1})$ and $\eta_2 = \tau_{nr2}/(\tau_{nr2} + \tau_{r2})$. $\tau_{tr}$ shown in Figure 2f is a prefactor to the Arrhenius term for the thermally activated transfer of excitons from population $n_2$ to $n_1$. The almost constant $\tau_{tr}$ around 80 ps confirms the assumed simple Arrhenius behavior with a fitted activation energy of $\Delta\varepsilon_1$ = 52 meV. A possible back transfer of population from $n_1$ to $n_2$ cannot be distinguished in our data due to the short lifetime of $n_2$.

To deconvolve the spectral contribution of $n_1$ and $n_2$, we fitted the time-dependent fluorescence intensity $I(E, t)$ at each wavelength as:

$$I(E,T,t) = \beta_1(T,E)\frac{dn_1(t)}{dt} + \beta_2(T,E)\frac{dn_2(t)}{dt} \qquad (2)$$

where $\beta_1(T,E) = \alpha\eta_1(T)A_1(E)$ and $\beta_2 = \alpha\eta_2(T)A_2(E)$. Since all decay processes are exponential, the derivatives in Eq. (2) are proportional to the respective populations. Each recombining exciton emits a photon with a probability $\eta_1(T)$ and $\eta_2(T)$ respectively, corresponding to the temperature-dependent PL quantum yields of the two populations. Each emitted photon triggers a count on the detector with a probability $\alpha$, which depends on the geometry of the sample and the measuring instrument and is assumed constant throughout all measurements. $A_1(E)$ and $A_2(E)$ are dimensionless functions whose integral over the whole spectral range is normalized to 1 and that reflects the shape of the PL spectra.

The fluorescence spectra of both populations $\eta_1(T)A_1(E)$ and $\eta_2(T)A_2(E)$, shown in Figures 3a-c are similar to single Gaussians with a width $\sigma$ of around 100 meV. If we observed a simple relaxation or exciton migration within an energy distribution of recombination centers, this would always result in a redshift for increasing time *t*. In such a situation, our model would always yield $A_1(E)$ red-shifted relative to $A_2(E)$. However, $A_1(E)$, which is the spectrum of the longer-lived exciton species, is red-shifted relative to $A_2(E)$ at 100 K and 323 K, while at 200 K it is blue-shifted. This confirms that we indeed observe two distinct populations of emitters with different temperature-dependent spectra.

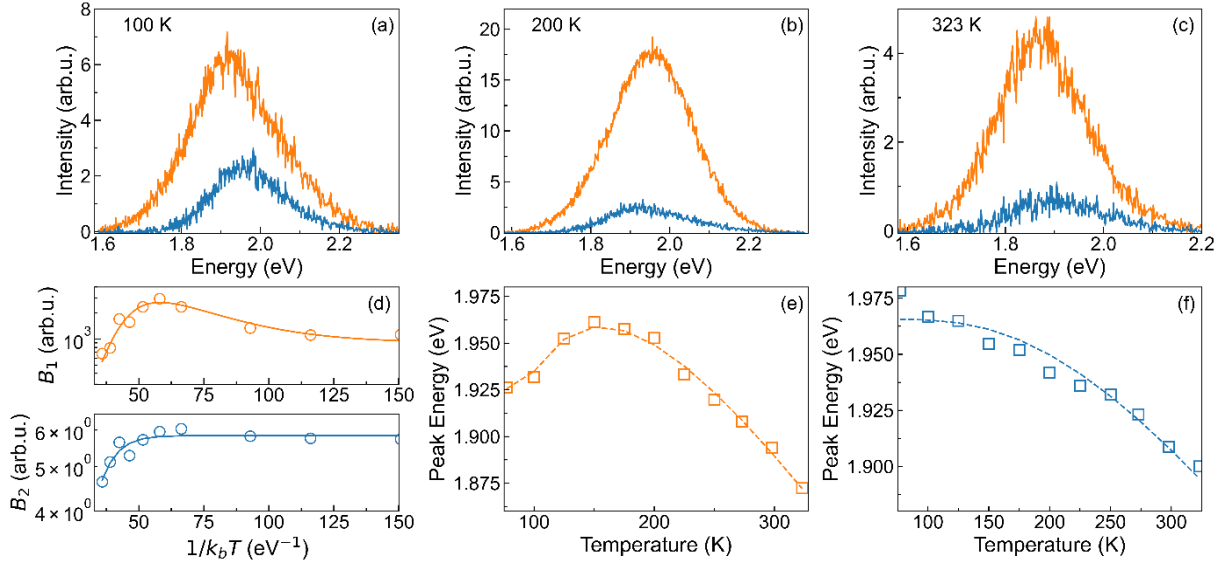

*Figure 3 a-c) fluorescence spectra of long- (orange curve) and short-lived (blue curve) excitons at 100K, 200K, and 323K, respectively; d) Arrhenius plot of the integrated intensities $B_1$ of long-(top panel, orange circles) together with the fit to Eq. (3) (solid orange curve) and $B_2$ of short-lived (bottom panel, blue circles) together with the fit to Eq. (4) (solid blue curve). e) and f) shift as a function of the temperature of long-lived centers (orange squares) together with the fit to Eq. (6) (dashed orange curve) and the short-lived centers (blue squares) together with the fit to Eq. (5) (dashed blue curve), respectively.*

Given that the fluorescence spectra of the two populations as plotted in Figures 3c are proportional to the PL quantum yields, we plot $B_1(T) = \int \beta_1(T,E)dE$ and $B_2(T) = \int \beta_2(T,E)dE$ in Figure 3d to reveal additional non-radiative recombination channels. Remarkably, $B_2(T)$ suggests an Arrhenius-like thermally activated non-radiative channel:

$$B_2(T) = \frac{B_2(0)}{1+\frac{\tau_2}{\tau_{nr2}}e^{-\frac{\Delta\varepsilon_{nr2}}{k_bT}}} \qquad (3)$$

where $B_2(0)$ is proportional to the quantum yield at 0K, $\tau_{nr2}$ is the Arrhenius prefactor of the additional thermally activated non-radiative channel and $\Delta\varepsilon_{nr2}$ is the activation energy.

Fitting our results with Eq. (3), we obtain $\tau_{nr2} \sim 0.5$ ps and $\Delta\varepsilon_{nr2} = 110$ meV.

$B_1(T)$, on the other hand, starts to increase from 125 K, reaches its maximum value at 200K, and then is quenched at higher temperatures. $n_1$ is populated predominantly from $n_2$ via thermal activation. Together with the introduction of an Arrhenius-type non-radiative term, this results in a more complex temperature-dependent quantum yield:

$$B_1(T) = \frac{B_1(0)}{\left(1+a_1 e^{-\Delta\varepsilon_{nr1}/k_b T}\right)^2}\left[1 + c\frac{a_2 e^{-\Delta\varepsilon_1/k_b T}}{\left(1+a_2 e^{-\Delta\varepsilon_1/k_b T}\right)}\right] \quad (4)$$

where $B_1(0)$ is proportional to the quantum yield at 0K, $\Delta\varepsilon_1$ and $\Delta\varepsilon_{nr1}$ are the activation energies of the transfer from $n_2$ to $n_1$ and the nonradiative processes for $n_1$, respectively, $a_1=\tau_1/\tau_{nr1}$, $a_2=\tau_2/\tau_{tr}$, and $c$ is the ratio between the initial populations $n_2$ and $n_1$.

This model is adapted from the one developed for a system including two emitter populations where (i) the carrier can recombine to the ground state from each of the populations; (ii) the species can migrate only from one population to the other and not vice-versa [29]. We note that this formalism has been developed for CW PL data. Since the CW PL intensity is proportional to the PL quantum yield, we can apply the same formalism to model our $B_1(T)$.

From the fit, we obtain $\Delta\varepsilon_1 \sim 45$ meV, which is in good agreement with the value of 52 meV obtained from the fit of the spectrally integrated dynamics with Eq. (1). For the non-radiative processes, we extracted $\tau_{nr1}\sim 2$ps and an activation energy $\Delta\varepsilon_{nr1}\sim 140$ meV. The $a_2$ ratio between $\tau_2$ and $\tau_{tr}$ extracted from the fit is $\sim 5$, which is higher than the one obtained from the fit of the dynamics to Eq. (1) without this additional non-radiative process ($a_2\sim 2$), while for $c$ we obtained $n_1$=12% and $n_2$=88%, in good agreement with the initial populations used to solve the rate equations (20%, 80%). The agreement between the fit parameters could be improved by iterating through Eqs. (1)-(4), but no added understanding would be gained. The activation energies of non-radiative processes of both $n_1$ and $n_2$ are ca. 3 times that of the transfer from $n_2$ to $n_1$. It is plausible that this activation involves either excitation into different bands within the crowded band structure of GeCH$_3$ [17], from which non-radiative recombination occurs, or thermally activated exciton dissociation. In the latter case, the activation energy would be a measure of the exciton binding energy.

For further insight into the nature of the two exciton populations, we fit $A_1(E)$ and $A_2(E)$ with a Gaussian curve and plot the peak positions as a function of the temperature in Figure 3e-f. The $A_2(E)$ peak position follows the O'Donnell and Chen model [30], which is a refinement of the empirical Varshni equation [31] and provides more insight into the electron-phonon coupling at the origin of the temperature-dependent bandgap:

$$E_{pl} = E_0 - S\langle E_{ph}\rangle \left[\coth\left(\frac{\langle E_{ph}\rangle}{2k_b T}\right) - 1\right] \quad (5)$$

where $E_0$ is the PL peak at 0K, $S$ is the Huang-Rhys parameter, and $\langle E_{ph}\rangle$ is the average energy of phonons coupling to the involved electronic levels. From the fit we obtain $E_0$=1.97 eV, $S$=5.3 and $\langle E_{ph}\rangle$= 66 meV.

The temperature-dependent peak position of $A_1(E)$ exhibits first a blueshift from 77K to 150K and a subsequent redshift from 175K to 323K. Such "S shape" behavior has been previously reported for the high energy band of the GeCH$_3$ PL emission [22], but it was not investigated in detail. The initial blue-shift followed by a red-shift of the emission peak energy has been observed in both CW and TRPL measurements for different semiconductor systems [29, 32-39], including excitonic materials with a certain amount of disorder, such as organic-inorganic lead-halide perovskites [40, 41], 2D transition metal dichalcogenides [42-44], or phosphorene [45]. It has been ascribed to thermal redistribution of excitons within an ensemble of width σ of localization centers with a mean activation energy ΔE. We obtain the temperature-dependent PL peak position [36]:

$$E_{pl} = E_0 - S\langle E_{ph}\rangle \left[\coth\left(\frac{\langle E_{ph}\rangle}{2k_b T}\right) - 1\right] - x(T)k_b T \quad (6)$$

$$xe^x = \tau_1/\tau_{tr} \left[\left(\frac{\sigma}{k_bT}\right)^2 - x\right] e^{\Delta E/(k_bT)} \qquad (7)$$

where the first two terms of Eq. (6) are the same as in Eq. (5). The third term accounts for exciton migration between RCs [35], with $x(T)$ being the solution of Eq. (7). $\tau_1$ and $\tau_{tr}$ are temperature-dependent and are taken from Figures 3d and 3f. From the fit reported in Figure 3f we extracted $E_0 = 2.1$ eV, $S = 5.2$, $<E_{ph}> = 30$ meV, $\Delta E = 37$ meV and, $\sigma = 145$ meV. Hence, we can conclude that the width of the distribution is similar to the width of the PL peaks and the activation energy for exciton migration is a fraction of this width.

Our results allow us to extract vital information about the nature of the two types of excitons at the origin of the observed fluorescence. The formalism of Eqs. (1), (3)-(7) has been developed for arrays of two species of quantum wells (QWs) [29, 32 - 39], where both species of QWs show a certain distribution of exciton energy. Analogously, germanane provides a disordered energy landscape for exciton migration. The temperature-dependent $A_1(E)$ and $A_2(E)$ peak positions suggest that exciton migration is relevant only for the longer-lived exciton species.

Intriguingly, the temperature dependence of $A_1(E)$ and $A_2(E)$ as described in Eqs. (5)-(7) arises from coupling to phonons with mean energy $<E_{ph}> = 30$ meV (240 cm$^{-1}$) for $A_1(E)$ and $<E_{ph}> = 66$ meV (530 cm$^{-1}$) for $A_2(E)$, suggesting that the two excitons preferentially interact with different vibrational modes of the lattice. The presence of the 1.97 eV emission after annealing suggests that the residual water bear sufficient concentration to change the electronic structure locally and to provide a high density of recombination centers that enables exciton migration between them. Quantum chemical calculations [17] have found a dense ensemble of electronic levels close to the valence and conduction band edges as a consequence of the small local structural distortions in each layer induced by the presence of $H_2O$. We can thus assume two types of emitting excitons that have their electrons and/or holes in different levels from this ensemble. Our results prescribe the following requirements for the two emitting states: (i) both have an allowed transition to the ground state but not towards the band edges, (ii) one of them can transition to the other via thermal activation, and (iii) they couple with different lattice modes. Concerning the transition from $n_2$ to $n_1$, the energy difference between the $A_1(E)$ and $A_2(E)$ peaks in Figures 3e and 3f varies strongly with temperature. This variation is inconsistent with a simple Arrhenius-like thermally stimulated transfer with a fixed activation energy of 37 meV, as assumed in Eq. (1) and confirmed in Figure 3f. An alternative mechanism has recently been proposed for the interlayer exciton recombination in a heterostructure of two 2D monolayers [46]. Low energy phonons periodically modulate the band structure between a direct and an indirect gap, leading to a recombination rate that has a temperature dependence very similar to an Arrhenius behavior with formal activation energy much higher than the energies of the phonons involved. We can assume a similar mechanism for the population transfer from $n_2$ to $n_1$ via a phonon-induced modulation of the band structure.

To summarize, we used TRPL spectroscopy at different temperatures to study the above bandgap fluorescence of $GeCH_3$ samples. We find two distinct populations of emitting excitons localized at RCs within the intercalated water. Compared to 2D transition metal dichalcogenides, research on Xenes and Xanes is still its infancy and still needs to investigate the exciton binding energy, exciton transport mechanisms, trions, biexcitons, and higher many-body effects as well as fluorescence quantum yield, and charge separation at interfaces as the groundwork for future applications in optoelectronics, light-harvesting, and sensing.

## Data Availability Statement

The datasets generated during and/or analyzed during the current study are available from the corresponding author on reasonable request.

## Author Contributions

E. C., J. E. G., C.D.A., and C.G. conceived the experiment. W. L. B. H. prepared the samples. S.S. performed the measurement. E.C., C.G., and S.S. analyzed the data. All the authors discussed the results and wrote the manuscript.

## Acknowledgments

E.C. and CV acknowledge financial support from MIUR PRIN aSTAR (Grant No. 2017RKWTMY), and from European Union's Horizon 2020 research and innovation program through the MSCA-ITN SMART-X (GA 860553). Sample synthesis by W. L. B. H. and J.E.G. was supported by the Center for Emergent Materials, an NSF MRSEC, under award number DMR-2011876.